\documentclass[a4,12pt]{article}
\usepackage[2cell,arrow,curve,matrix,web]{xy}
\UseHalfTwocells \UseTwocells
\usepackage{graphicx}
\begin{document}

\begin{titlepage}

%\begin{flushright}

%q-bio/0410019

%\end{flushright}

\vspace{.5cm}

\begin{center}

{\Large \bf  Some Theoretical Investigations in EEG Studies}

\vspace{1cm}

%\footnote{indranil@theory.saha.ernet.in},
{\large I. Mitra{{\footnote{Permanent Address: Department of
Physics, New Alipore College, L Block, Kolkata 700 053, India}}
\footnote{imitra@ictp.it, mindranil00@yahoo.com}}$^a$}
%R. Chakraborty{\footnote{indranil@theory.saha.ernet.in}},$^b$\,}
\vspace{5mm}

{\em $^a$ Abdus Salam ICTP, Strada Costiera 11,\\
34014 Trieste, Italy}

%{\em $^b$Department of Physics, New Alipore College\\
%L Block, Kolkata 700 053, India}\\
%{\footnote{indranil@theory.saha.ernet.in}}\\
%{\em $^b$ Rashmandir Vidyalaya,\\
% BK -61,Salt Lake City, Kolkata-91}\\
%{\footnote{mitrarc@yahoo.com}}\\

\vspace{.5cm}

\end{center}

\vspace{.5cm}

\centerline{{\bf{Abstract}}}

\vspace{.5cm}

\begin{small}
It is well known that Electroencephalography(EEG) and the
respective evoked potentials have deep implications corresponding
to specific cognitive tasks and in the diagnosis of several
diseases such as epilepsy and schizophrenia. Some recent
experimental results have already shown some evidence of chaotic
activity in the brain. The Hodgekin-Huxley(HH) models, may yield
geometrical solutions in terms of limit cycles and basins of
attractors, but its implementation requires a priori knowledge of
the kinetics of the innumerable conductances acting in a given set
of cells. We are of the opinion that the EEG data should reflect
the neuronal dynamics, and there should be some mechanism at the
neuronal level which generates stochasticity compatible with the
recorded data. In this paper we develop a theoretical framework to
show that EEG dynamics may be governed by a suitably biased
Vander-Pol oscillator which is closely related with the modified
version of the FitzHugh-Nagumo(FN) model making extension of the
ideas of dynamic causal modelling (DCM). Eventually we also give a
prescription to compute the correlation matrices which may be
tested empirically, for some small values of the parameters.

\end{small}

\end{titlepage}
\newpage

\section{Introduction}
That the human brain is a complex system with significant
spatiotemporal dynamics is beyond doubt. EEG, a noninvasive
technique for probing the dynamics of brain provides a direct
measure of cortical activity with millisecond temporal resolution.
It should be mentioned here that EEG data is a set of changes in
voltage potential {\cite{a1,a01}} which is generated when neuronal
activities are activated. From the functional point of view EEG
studies have been engaged in a variety of brain activities
including cognitive tasks, showing the memory contents which are
activated showing the information processing operations, but most
abundantly, or to say clinically the research has remained
confined {\cite{a2}} with the analysis of several epileptic and
schizophrenic patients with comparative EEG analysis. Recently
some studies have also been initiated {\cite{a3}} in analyzing the
emotional states of patients by studying brain electric fields
from the EEG data.

For the analysis of EEG data, representations based on a Fourier
transform have been most commonly applied. These methods have
proved to be effective for various EEG characterizations, but fast
Fourier transform (FFT), suffer from large noise sensitivity.
Another approach based on neural network detection systems have
been proposed {\cite{a4,a5}} but with a false detection rate.
Interestingly {\cite{a6}} have evaluated different parametric
models on a fairly large database of EEG segments. Using inverse
filtering, white noise tests, and 1-s EEG segments, they found
that autoregressive (AR) models of orders between 2 and 32 yielded
the best EEG estimation. Recent techniques include methods
measuring chaos theory and the energy measure. The measure of
complexity used in chaos include parameters like correlation
dimension, similarity, synchrony and Lyapunov exponent. It has
been overemphasized over the years to find global solutions to
cable equations which represent long wavelength EEG standing
waves. It should be mentioned here that a nonlinear component is
apparent in all analyzed EEG records {\cite{a7,a70,a71}}. The time
series obtained by theoretical investigations {\cite{a8}} belonged
to four classes, suggesting a chaotic activity with correlation of
5.5. EEG was characterized by inspiratory bursts of oscillations
that disappeared during expiration and simulations suggested that
this state corresponds to a limit cycle attractor that is specific
to a given stimulus.

EEG's represent the integral output of a large number of neurons,
with a complicated dynamics of subsystems with innumerable degrees
of freedom. In addition the presence of noise of unknown origin
makes it hopeless to reinterpret the data within the framework of
chaos theory. Despite this difficulties epilepsy, one of the
cherished arenas of investigation of EEG remained to be a
recognized model of neuronal synchronization and it is now widely
believed {\cite{a9}} that seizure episodes are characterized by
bifurcations to system states of low complexity. To cite an
example, epileptic bursts produced in CA3 region of the rat
hippocampal slices are exposed to $K^{+}$ enriched extracellular
medium by electrical simulation of the mossy fiber inputs
{\cite{a10}}. Time series of the evoked field potentials were
analyzed and the conclusion was that of undoubted evidence for
chaos. Dimensional analysis of an epileptic EEG {\cite{a11}} again
provided a hypothesis of an existence of chaotic attractor being
the direct consequence of the deterministic nature of brain
activity. So by now it can be well inferred that EEG recordings
which are the marker for the neuronal activity which show many
characteristics of chaotic activity.

Taking all these into consideration we in this paper are mainly
concerned to show whether there exist any mathematical model to
fit into these scheme of things in explaining EEG phenomena. It
should be admitted here that we have not done any numerical
simulation, but mainly followed a mathematical formalism to show
whether the desired response of the EEG phenomena may be generated
using this mathematical structure. We have tried to investigate
the FitzHugh-Nagumo(FN) model though not in the original form. The
FN model is based on the premise that changes in the membrane
potential is related to sodium activation, inactivation and
potassium activation. It is a two variable model, but in essence
with our early comments this model unlike HH models is unable to
generate chaos. Though it should be mentioned here that some
modifications to the FN model has already been done giving rise to
Hindmarsh-Rose (HR) model, which is essentially a 3 dimensional
model which has the ability to generate chaos {\cite{a12}}.

The non linearity of the active membrane, the brain's high degree
inhomogeneity makes the electric field in the brain difficultly
complicated. Our approach is mainly based on the fact that EEG
data is measured as the difference in electric potential between
one and more other electrodes given by
\begin{equation}\label{ep}
    \Phi(\vec{r},t) = \frac{1}{4\pi\sigma}\Sigma_{i=1}^{n}\frac{I_{i}(t)}{R_{i}}
\end{equation}
where $I_{i}(t)$ is the fluctuating current moving from the i'th
current source into a medium of conductivity $\sigma$. $R_{i}$ is
the distance of the i'th source from the field point $\vec{r}$.
The potential difference  $V$ across the membrane follows the
diffusion equation given by
\begin{eqnarray}\label{memdiff}
    \lambda^{2}\frac{\partial^{2}\Phi}{\partial r^{2}} - \tau \frac{\partial \Phi}{\partial
    t} - \Phi &=& J
\end{eqnarray}
where $\lambda$ is the space constant of the axon, which is
determined by the electrical and geometrical properties of the
axon and the surrounding nerve cells, $\tau$ is the time constant
of the membrane and $J$ is the current source term. Some
experiments have already been done for some simple systems which
show that if we take, say two neurons with both excitatory and
inhibitory synaptic activity the potential measured at an
intermediate position ( assuming dipole currents with fluctuating
frequencies $f_{a},f_{b}$) and the temporal component can be
written as
\begin{eqnarray}\label{epsp}
\Phi(t) &=& \frac{1}{4\pi\sigma}[\frac{I_{a}(t)}{R_{1}}\cos(2\pi
f_{a}t + \alpha_{a}) -
 \frac{I_{a}(t)}{R_{2}}\cos(2\pi f_{a}t + \alpha_{a}) \nonumber \\
 &+& \frac{I_{b}(t)}{R_{3}}\cos(2\pi f_{b}t + \alpha_{b}) - \frac{I_{b}(t)}{R_{4}}\cos(2\pi f_{b}t +
 \alpha_{b})]
\end{eqnarray}
So essentially we see that EEG phenomena {\cite{a13}} is a form of
some disturbance, or mainly a wave phenomena which may or may not
be stochastic in nature. The reason which motivated us to write an
article on the analysis of EEG is crucially based on the
observation that the EEG potential may be regarded as a function
of surface coordinates, which may give rise to stochasticity due
to the intrinsic network dynamics and may be represented as a
superposition of a number of travelling waves, given by
\begin{eqnarray}\label{trav}
    \Psi(t, x, y) &=& \sum_{l}\sum_{m}\sum_{n}C_{lmn}sin(2\pi
    f_{n}t - k_{xl}x - k_{ym}y)
\end{eqnarray}
which can be argued to be analogous to Eq.~(\ref{epsp}) if we
include the spatial degrees of freedom.

So the goal of this paper is to identify an underlying scheme of
events which may play a crucial role in determining the EEG
patterns as depicted above and we will try to show how
nonlinearity may be fitted in those models. In the next section we
discuss about some general considerations on EEG and a brief
review of the existing models. In section \ref{vpes} we show the
analogy of a simple nonlinear oscillator, the biased Vander-Pol,
with some extra parameters, which is proposed to be a good
candidate as the FN model of neuronal behavior and show with the
help of Langevin formalism, that an equivalent Fokker-Plank (FP)
equation corresponding to it may be developed and show how to
define the EEG potential through the probability distribution and
define some measurable quantities which may have the required
stochastic response to be fitted into the recently made
experiments of EEG data.
\section{Some notes on EEG and it's extensions}

As we have already stated in the introduction in general all EEG
phenomena have a temporal behavior with different categories. A
spectral analysis may be doe on any one of the categories. To be a
bit quantitative a sinusoidal function $\Psi(t)$ is completely
described by three parameters as $ \Psi(t) = C \sin(2\pi f t +
\phi)$ and in general the epoch will be a superposition of the
form
$$ \Psi(t) = \Sigma _{n}^{k}C_{n} \sin(2\pi f_{n} t + \phi_{n})$$
The spectral density function $Q(f)$, a measure of the $C_{n}^{2}$
is obtained as a Fast Fourier Transform(FFT) with the following
prescription
\begin{eqnarray}\label{spec}
Q(f) &=& \frac{2\Delta t}{N}{{\|\sum_{l = 0}^{N -
1}\Psi_{l}exp(-i2\pi lf\Delta t)\|}}^{2}
\end{eqnarray}
where $\Psi_{l}$ refers to the EEG sampled at times given at
discrete time intervals $l$. Similarly we can define the cross
spectral density functions for two channels of EEG,
\begin{eqnarray}
Q(f) = \frac{2\Delta t}{N}[\sum_{l = 0}^{N - 1}\Psi_{1l}exp(-i2\pi
lf\Delta) \sum_{l = 0}^{N - 1}\Psi_{2l}exp(-i2\pi lf\Delta)]
\end{eqnarray}
 to
get a coherence function which will give us a measure of
correlation between the signals. Now if we include the spatial
coordinates where the basic epoch of EEG is given by
Eqn.~(\ref{trav}), then the covariance matrix consisting of $M$
epochs of EEG from the cross spectral density functions
Eqn.~(\ref{spec}) of each epoch may be written a
$$ q_{rs}(f) = \frac{1}{M}\sum_{l = 1}^{M}Q_{rs}^{l}(f)$$
where $r,s$ are as usual channel numbers and $l$ is the epoch
number. The frequency wavenumber spectrum, $T(f, k_{x}, k_{y})$ a
measure of $C_{lmn}$ can be obtained from the covariance matrix.
So the basic scheme is that simple, analyze the EEG for say two
cases, one normal and another epileptic, find out the spectral
density in both cases, analyze the results, make a statistical
inference and go out clinically. But the actual point here is that
as far as the comparison is concerned at a particular scale it may
be fine, but does the analysis take into account the real
dynamical features of the brain. In other words the basic epochs
which we have considered may be constructed assuming a basic
circuit model which generate a travelling wave solution as is the
case in Eqn.~(\ref{memdiff}). What we are more importantly
concerned over here is to take into account the recent
experimental results {\cite{a14,a15,a16}} and to propose an
underlying dynamics for EEG phenomena. Recent models of EEG
studies {\cite{a17}} has given place to fMRI analysis which are
executed as a general linear model
$$ y = X\beta + \eta$$
which measures the experimentally controlled investigations to the
observed blood oxygen level dependent responses (BOLD)
{\cite{a171}}, $y$ in a voxel specific fashion with the design
matrix $X$ and a gaussian noise $\eta$. There has been extensions
of this idea in terms of Multivariate autoregressive models(MAR)
{\cite{a18}}, modelling the vector of regional BOLD signals at
time $t(y_{t})$ as a linear combination of k past data vectors,
with an weighted contribution of the parameter matrices $A_{i}$
and a noise term $\eta$
\begin{eqnarray}\label{bold}
    y_{t} &=& \sum_{i=1}^{k}y_{t-i}A_{i} + \eta_{t}
\end{eqnarray}
The drawback of the MAR models is that they do not involve
biophysical forward terms, which enable inferences about neural
parameters. This model has been extended by DCM {\cite{a19,a20}},
which constructs a reasonably realistic neuronal model of
interacting cortical regions with neurophysiologically meaningful
parameters. In DCM, neural dynamics in several regions,
represented by a neural state vector $z$ are driven by
experimentally designed inputs. In DCM the change in neural states
is a non linear function of the states $(z)$, the inputs $(u)$,
and the neuronal parameters $(\theta^{n})$, which are the
connectivity matrices defining the functional architecture and
interactions among brain regions at a neuronal level. The general
equation may be written as
\begin{eqnarray}\label{dcm}
\dot{z} &=& F(z, u,\theta^{n} )
\end{eqnarray}
It should be noted here that the DCM model may be extended to
include background noise which is correlated in space and time. We
emphasize this point as we will proceed with this formalism in our
case. It may be assumed that the recorded signal is a simple
superposition of the brain response and the background noise which
is the signal plus noise (SPN) {\cite{a21}} model, where the
measured signal $\Lambda_{rs}^{t}$ at channel $r$ and time sample
$s$ in trial $t$ is formulated as $\Lambda_{rs}^{t} = \Lambda_{rs}
+ \eta_{rs}^{t}$, where $\Lambda_{rs}$ is the brain response
caused by the stimulus and $\eta_{rs}^{t}$ is the measured noise,
which is correlated in space and time as a covariance matrix
$\Sigma = X \otimes T$, where $X, T$ are spatial and temporal
covariance matrices respectively.

Some recent theories in interpreting the set of images in terms of
visual signals corresponding to brain activities may be given by a
Markov Random Field (MRF) {\cite{a22}}, implying that probability
of a pixel, assuming a particular value is dependent on the values
of the neighbouring pixels. It is governed by a joint probability
distribution
\begin{equation}\label{mrf}
p(x) = \frac{1}{Z}exp(-\beta U(x))
\end{equation}
where $Z$ is the normalization constant, $\beta$ the inverse of
temperature, $x$ is a vector denoting an array of pixel values and
$U(x)$, the Potential. Once the choice of the potential is made
the method of simulated annealing (SA) {\cite{a23}} may be used to
employ a stochastic differential equation, which gives the
dynamics of the image formation
\begin{eqnarray}\label{ham}
    dy(t) &=& -\nabla U(y) + \sqrt{\frac{2}{\beta}}dW
\end{eqnarray}
where $y$ denotes the image, $U$ the potential and $W$ the
Brownian process. Now we have some points to make regarding the
Eqn.~(\ref{ham}) which will be pertinent for our purpose. The
equation is essentially the It$\hat{o}$ interpretation
{\cite{a24}} of the standard Langevin equation
\begin{eqnarray}\label{lang}
\dot{y} &=& A(y) + D(y)L(t)
\end{eqnarray}
We will be considering equations of the form (\ref{ham}) and to
avoid the IS dilemma we will be interpreting the noise as an
external one which is essentially created in an otherwise
deterministic system (though may not be the case with brain, but
at least as of now we don't have a choice) by a random force whose
stochastic properties are assumed to be known.

\section{Biased Vander-Pol Equation as an aid to understand EEG}\label{vpes}

It is beyond doubt that a number of cellular and combined cellular
network mechanisms generate collective behavior of neurons. Cells
which may not be intrinsically oscillatory may become so as a
consequence of network properties {\cite{a25}}. As there are
multiple interacting levels of organizational hierarchy in the
human brain there are suitable forms of cooperativity and
synchronization. It is a possible gesture that the mean field
dynamics {\cite{a26}} may give us an useful insight into the
proper realm of things. So in our objective for a proper
understanding of the dynamics which make the brain visible through
EEG, which is the analysis of brain waves we first of all
postulate that that the EEG is represented by the following
expansion
\begin{eqnarray}\label{eeg}
 \Psi(t, x, y) &=&
 \sum_{n}\sum_{l}^{M}C_{nl}E[H_{nl}(x,y)]e^{(i2\pi f_{n}t)}
\end{eqnarray}
Here ${H_{nl}}(x,y)$ are the empirical orthogonal functions which
reflect the spatial properties corresponding to the n'th temporal
frequency component of the EEG and $M$ denotes the number of EEG
channels. It should be mentioned here that $E[H_{nl}(x,y)]$ is the
expectation value of the functions with respect to a particular
distribution function. As we will see shortly that the
distribution function arises as a solution to a stochastic
differential equation which as we understand should be governed by
the underlying cortical dynamics. The functions ${H_{nl}}(x,y)$
are the eigenvectors of the covariance matrix $ q_{rs}(f_{n})$ and
the coefficients $C_{nl}$ can be related to it's eigenvalues.
Again the estimate for the frequency-wavenumber spectrum $T(f,
k_{x}, k_{y})$may be obtained by giving the following definition
\begin{eqnarray}\label{fwspec}
T(f, k_{x}, k_{y}) = \frac{1}{M^{2}}\sum_{r = 1}^{M}\sum_{s =
1}^{M}q_{rs}(f)E[H_{nl}(x,y)]
\end{eqnarray}
So essentially in understanding and interpreting the electric
fields or EEG data Eqn.~(\ref{fwspec}) tells us all. The
covariance matrix is the matrix formed by the cross spectral
density functions and contains complex numbers as they contain the
relative phase of EEG. So this part is purely a part of the
experimental data, but the other part is dependent on the
underlying mechanics and cortical activity. The better the
guesswork made in devising a cortical model we will be able to get
a better fit.

Now to analyze the dynamics of the cortical network which may give
rise to oscillations we consider a particular generalization of
the Van der Pol (VP)oscillator. It is well known that VP models
are special case of the Lienard equations $$ \ddot{x} + f(x)
\dot{x} + g(x) = 0$$ which with some stringent conditions
{\cite{a28}} on the functions $f, g$ may lead to unique, stable
limit cycle surrounding the origin in the phase plane. The VP
equation in it's original form $$ \ddot{x} + \mu(x^{2} -1)\dot{x}
+ x = 0$$ has a degenerate Hopf bifurcation at $\mu =0$, which
implies the vanishing of the nonlinear term, but if we scale the
variables as $ z = \mu^{1/2}x$ we may be able to get a Hopf
bifurcation without the vanishing of the nonlinear term. So it may
be anticipated that the VP models may be used to study bursting
behavior of the neurons, which eventually leads to evoked spike
trains which appears in phase space to undergo a transition from a
steady state to a repetitive limit cycle via Hopf bifurcations
{\cite{a29}}. So we extend the biased VP model as
\begin{eqnarray}\label{vp}
\ddot{x} + \mu(x^{2} - 1)\dot{x} + \nu Z(x) &=& a + \Lambda(t)
\end{eqnarray}
Here the parameters $\mu,\nu, a$ though independent of
spatio-temporal variables, yet may depend on some other variables
linked with the cortical architecture and function( for example if
we take the neurons as point particles, they may be related to
their field, mass etc), $Z(x)$ a choice function and $\Lambda(t)$
is a noise term. Now it is important to mention here that the
above model may be implemented as a equivalent circuit with
several components and storage devices which may be a sufficient
criterion to generate chaos, by the Poincare-Bendixon theorem
{\cite{a24}}. Now as has been already advocated {\cite{a31}} that
Eqn.~(\ref{vp}) can be related to a modified FN model which by the
previous argument may give rise to chaotic behavior. It can be
shown by the Lienard plane analysis that when $\mu >> 1$ the
system has a stable limit cycle for a critical value of the
parameter $a$, which show relaxation oscillations phenomena. But
it should be be mentioned here that in the modified version of our
VP model, for some suitable choice function we do propose the
possibility of bifurcations. In this context to gain some more
grounds into the proposal, we can't resist the temptation of
introducing another oscillator, a special case of the Forced
Duffing (FD) oscillator with weak parameters as
\begin{equation}\label{fd}
    \ddot{x} + \dot{x} + x + \epsilon x^{3} = F(t)
\end{equation}
The above system may be interpreted as a weak perturbations of the
harmonic oscillator and gives rise to saddle node bifurcations of
cycles. But unfortunately we don't find any relaxation processes
associated with it and thereby could't be guaranteed that it may
serve as a viable neuronal architecture model. With this
digression we would like to visit our modified VP model
Eqn.~(\ref{vp}). As in our formalism we have to compute the
expectation values of the empirical orthogonal functions we  would
like to recast Eqn.~(\ref{vp}) in the Langevin approach where we
get an equivalent two dimensional system given by
\begin{eqnarray}\label{vpl}
    \dot{x} &=& f(x) + y \nonumber \\
\dot{y} &=& -\gamma y + \epsilon \Omega(x, \epsilon y) +
\Lambda(t) + a
\end{eqnarray}
with $f(x) = (\gamma + \mu) x - \frac{\mu x^{3}}{3}$. It should be
noted here that $\gamma$ is defined in terms of the given
parameters in the modified VP equation and the $\Omega$ is
determined in terms of $ f(x)$ and $Z(x)$, though an arbitrary
$\epsilon$ dependence may not be removed. Here we take $\epsilon
<< 1$ which ensures the equivalence of the two systems up to
$O(\epsilon^{2})$. We assume a non gaussian noise term, to have a
viable chaotic interpretations and it is assumed to have the
following properties
\begin{eqnarray}\label{noi}
   <\Lambda(t)> &=& 0 \nonumber \\
<\Lambda(t_{1})\Lambda(t_{2})> &=&  \vartheta \delta(t_{1} - t_{2})\nonumber \\
<\Lambda(t_{1})\Lambda(t_{2}) \cdots \Lambda(t_{m})> &=&
\vartheta_{m} \delta(t_{1} - t_{2})\delta(t_{1} -
t_{3})\cdots\delta(t_{1} - t_{m})
\end{eqnarray}
for $m\geq 1$. It should be mentioned here that $\vartheta_{m}$
for $ m = 2$ determines the size of the fluctuating term and
related to the thermal noise {\cite{a34}} in the problem. So
essentially the noise term is essentially determined by the
constants $\vartheta_{m}$. To get the master equation for this
system, if we consider a Poisson process $Z(t)$ given by $ Z(t) =
\int_{0}^{t}\Lambda(t')dt', (t > 0)$, which is characterized by
the transition probability $T_{\tau}(z)$ defined as the product of
convolutions as
\begin{eqnarray}\label{tran}
\int{e^{ikz}T_{\tau}(z)dz} &=& exp[\tau\sum_{m =
1}^{\infty}\frac{(ik)^{m}}{m!}\vartheta_{m}] =
exp[\rho\tau\int{(e^{ikz} - 1) w(z) dz}]
\end{eqnarray}
with $ \vartheta_{m} = \rho\int{z^{m}w(z)dz}$.

Now with all these definitions and some suitable restrictions on
the noise function it could be seen that Eqn.~(\ref{vpl}) is
equivalent to the FP equation given by
\begin{eqnarray}\label{fp}
\frac{\partial P(x,y,t)}{\partial t} &=& -y\frac{\partial f(x)
P}{\partial x} + \frac{\partial(\gamma y + \epsilon\Omega x)
P}{\partial y} + \rho\int{w(\zeta)(P(x, y + \zeta x, t) - P(x, y,
t)d\zeta}
\end{eqnarray}
with some suitable initial conditions on the Probability
distribution dictated by the linearized version of the DCM
Eqn.~(\ref{dcm}). So essentially with some approximations we have
found out the stochastic differential equation determining the
probability distribution of the underlying relaxation oscillator
process driven by the modified VP equation, which may in some
sense characterize the neuronal processes. So the task which
remains in hand is to calculate the EEG and the
frequency-wavenumber spectrum which requires the knowledge of the
expectation values which is defined by
\begin{eqnarray}\label{exp}
    E[H(x(t), y(t))] &=& \int\int {H(x,y) P(x,y,t) dx dy}
\end{eqnarray}
It should be noted here that we may define the time correlation
matrix of any quantity $G(x,y)$ between any two time intervals by
$$\chi(t) =
\int_{t_{1}}^{t_{2}}{d\zeta \delta G(x(t + \zeta), y(t +
\zeta))\delta G(x(\zeta), y(\zeta))^{T}}$$ where $ \delta G = G -
E[G]$

It is to be noted here that the Langevin formalism in one
dimensional formulation with a color noise in the joint variables
$x, \xi$
\begin{eqnarray}\label{langc}
    \dot{x} &=& f(x) + \xi \nonumber \\
    \dot{\xi} &=& -\gamma \xi + L(t)
\end{eqnarray}
gives rise to a FP equation of the form
\begin{eqnarray}\label{fpls}
    \frac{\partial P}{\partial t} &=& \frac{\partial}{\partial
    x}f(x)P + \frac{1 - e^{-(1 + \gamma)t}}{1 + \gamma}\frac{\partial^{2}P}{\partial y^{2}}
\end{eqnarray}
Corresponding to this, it is worthwhile to mention that an action
functional has been proposed {\cite{a37,a38}} in some
input($i(x)$) output ($o(x)$) variables as
\begin{eqnarray}\label{act}
    S[o(x), i(x)] &=& j\int{dt  i(x)(o(x)\dot{x} - f(x))} +
    \vartheta\int{dt {i^{2}(x)}}
\end{eqnarray}
Expanding S in a Volterra expansion upto the second order may give
rise to a probability functional
\begin{eqnarray}\label{pfunc}
 P[o(x, t), i(x, t)] &\sim& e^{-\frac{1}{2}\int{d^{2}x d^{2}x' \int{ dt dt'o(x,t)(o(x',t')K(x, x', t, t')
 + i(x', t')M(x,
 x', t, t'))}}}
\end{eqnarray}
where $K, M$ are functions, which are dependent on the neuronal
architecture. This formalism has been applied to the Limulus eye,
{\cite{a39}} whose equation is already known to get hold of a
specific form of these functions. Though this is not the
probability distribution we were talking about, but it will be
interesting to see in this case that under what approximations the
distribution matches with the functional.

An important point which is to be noted here is that the modified
VP Eqn.~(\ref{vp}) which we have used here is a complicated
equation and in general the solution is in many instances
difficult to determine. But if the parameters are small we may
bypass the solution to find out the expectation values by
perturbation theory. To give an example in the case of the FD
oscillator Eqn.~(\ref{vp}), with F(t) interpreted as white noise
the FP equation is given by
\begin{eqnarray}\label{fpfd}
\frac{\partial P}{\partial t} &=& -\dot{x}\frac{\partial
P}{\partial x} + \frac{\partial(\dot{x} + x + \epsilon x^{3})
P}{\partial \dot{x}} + \frac{\partial^{2}P}{\partial \dot{x^{2}}}
\end{eqnarray}
The equation for the expectation values show that a complete
analytical solution is possible when $\epsilon$ is small and we
can treat it perturbatively.

\section{Conclusions}
We have tried in this article to give a theoretical framework to
model EEG data which takes into account the underlying neuronal
dynamics. As the EEG data is essentially thought to be consisting
of travelling waves which is inherently nondeterministic, the
formalism is based on finding out an analysis based on a suitable
relaxation oscillator which has the potential to generate
stochasticity. We also give a theoretical scheme of how to compute
the EEG spectrum, the correlation and covariance matrices based on
the modified version of the FN model. One can make some numerical
estimates of the spectrum based on some fixed small values of the
parameters and an assumed Gaussian probability distribution in the
simplest case and see how it matches with the experimental data,
based on FFT which in turn may give some definite indications
about the correctness of the assumed model. So based on this
hypothesis we hope that experimental studies of EEG, apart from
clinical importance have a large role to play in brain modelling.
Recent experimental results indicate that wavelet transforms (WT)
{\cite{a41}} are much suited for EEG studies,which depends on the
scaling and shifting properties of the initial wavelet. It will be
interesting to explore the consequences and extrapolations of the
proposed theoretical framework in the context of the WT model.

\section*{Acknowledgements}
The author sincerely thanks the ICTP hospitality where the work
has been conceived and done.

\end{document}